# V-CARE: A Blockchain Based Framework for Secure Vehicle Health Record System

Pranav Kumar Singh, *Member, IEEE,* Roshan Singh, and Sukumar Nandi, *Senior Member, IEEE*

**Abstract**

One of the biggest challenges associated with connected and autonomous vehicles (CAVs) is to maintain and make use of vehicles health records (VHR). VHR can facilitate different entities to offer various services in a proactive, transparent, secure, reliable and in an efficient manner. The state-of-the-art solutions for maintaining the VHR are centralized in nature, mainly owned by manufacturer and authorized in-vehicle device developers. Owners, drivers, and other key service providers have limited accessibility and control to the VHR. We need to change the strategy from single or limited party access to multi-party access to VHR in a secured manner so that all stakeholders of intelligent transportation systems (ITS) can be benefited from this. Any unauthorized attempt to alter the data should also be prevented. Blockchain is one such potential candidate, which can facilitate the sharing of such data among different participating organizations and individuals. For example, owners, manufacturers, trusted third parties, road authorities, insurance companies, charging stations, and car selling ventures can access VHR stored on the blockchain in a permissioned, secured, and with a higher level of confidence. In this paper, a blockchain-based decentralized secure system for V-CARE is proposed to manage records in an interoperable framework that leads to improved ITS services in terms of safety, availability, reliability, efficiency, and maintenance. Insurance based on pay-how-you-drive (PHYD), and sale and purchase of used vehicles can also be made more transparent and reliable without compromising the confidentiality and security of sensitive data.

## I. Introduction

Modern vehicles have a complex mechatronic structure, which is not only a mechanical system of engine, brakes, gearbox, accelerator; now, it is like a networked computer on wheels (NCoW) with sophisticated control, sensing and communication layers and over millions of lines of code [1]. Fig. 1a illustrates the typical in-vehicle subsystem, which comprises sensors, actuators, Controller Area Network (CAN), Electronic Control Units (ECUs), GPS, Vehicle-to-Everything communication (V2X), On-Board Diagnostic (OBD) framework, Onboard Unit (OBU), central gateway, etc. [2].

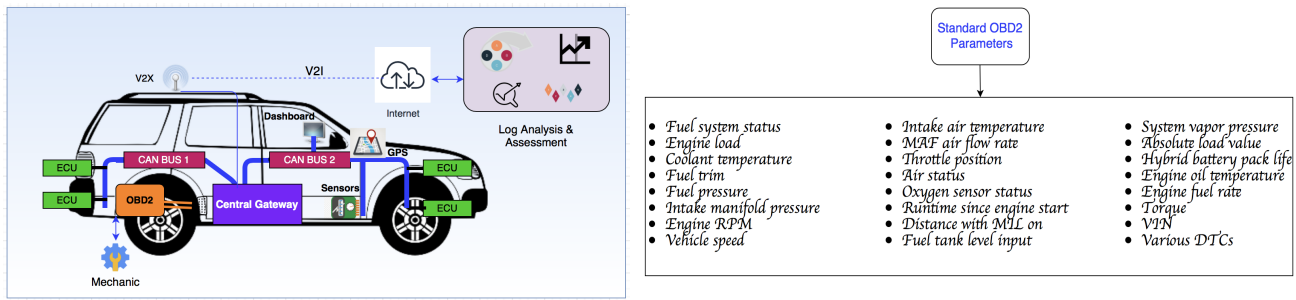

(a) VHR from In-Vehicle Network Architecture (b) Standard OBD2 Parameters

Fig. 1: Functional Diagram of Common VHR Monitoring and standard OBD2 Parameters

There are different protocols for data exchange between inter-subsystem (ECUs) and intra-subsystem (within each subsystem) such as CAN (powertrain sensors, transmission, engine controller), CAN with Flexible Data Rate (CAN-FD), FlexRay (Airbag, chassis, steering, brake control), Local Interconnect Network (LIN) (instrument cluster, door, seat, light, climate, seat, light, climate), Media Oriented Systems Transport (MOST) (phone, audio, display, navigation),

Pranav Kumar Singh is with the Department of Computer Science and Engineering, Indian Institute of Technology Guwahati, 781039 and Central Institute of Technology Kokrajhar, 783370 Assam, India (e-mail: snghpranav@gmail.com)

Roshan Singh is with Open Source Intelligence Lab, Department of Computer Science and Engineering, Indian Institute of Technology Guwahati, Assam 781039, India (e-mail: roshansingh3000@gmail.com)

Sukumar Nandi is with the Department of Computer Science and Engineering, Indian Institute of Technology Guwahati, Assam 781039, India (e-mail: sukumar@iitg.ac.in

and Ethernet. The data from different ECUs is accessible via the central gateway to authorized persons. The OBD2 protocol specified in SAE J1979 can also log a range of vehicle data. Some of the standard OBD2 parameters are listed in Fig. 1b. OBD2 provides access to data that indicates the status of the various subsystems to the vehicle owner or the mechanic cable/Bluetooth/hotspot/cloud access. The raw data collected via OBD2 and gateway data is transmitted to the cloud through the vehicle-to-infrastructure (V2I) communication and stored as VHR, which is used for intelligent analysis (diagnostics) and assessments (prognostic) using APIs. Diagnostics is analysis of current state of vehicle subsystems whereas prognostic is assessment of the future state of vehicle subsystem [3]. The owners/drivers access these details on their smartphones/dashboards via apps provided by the cloud service providers (manufacturer/trusted third party).

*A. Problem Definition*

The state-of-the-art solutions for VHR are centralized, which is based on cloud-based solutions. The present system is dedicated only to diagnostic and prognostic purposes i.e., finding the existing faults and predicting the future faults. Though such vehicle health monitoring systems provide clear benefits to owners/drivers for the maintenance of their vehicles, the data from vehicles ECUs can create additional opportunities in terms of safety, services, insurance, EV-charging, sales and purchasing of used vehicles, analysis of driving behavior, etc. However, these value additions require a change in the traditional deployment strategies, and there are serious issues that need to be addressed such as security, privacy, data consistency, access control, performance, availability, reliability, level of transparency, etc.

*B. Related Work*

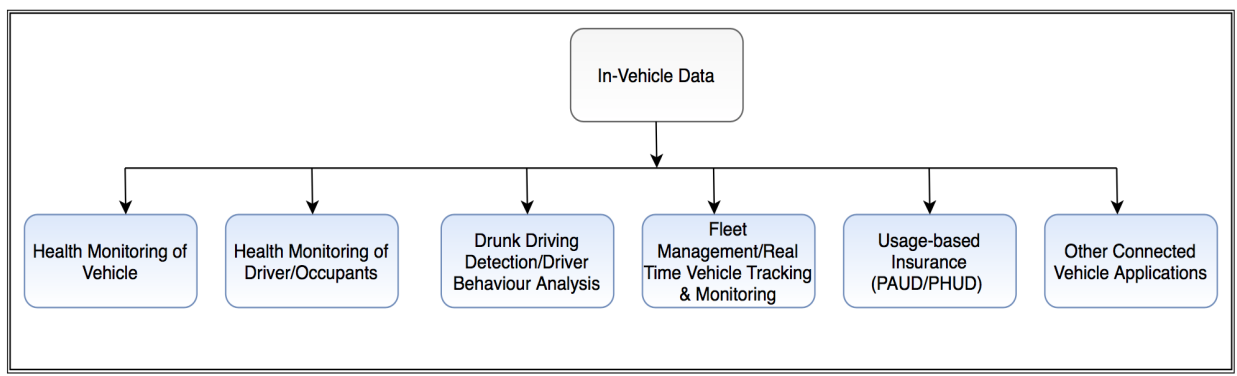

Fig. 2: In-Vehicle Data Applications

Several years of research on capitalizing the in-vehicle data (from sensors/ECUs) have resulted in various applications [3], [4], [5], [6], [7], [8], [9] as illustrated in Fig. 3. With the explosion of sensors and the revolution in telematic and radio access technologies (RATs), a normal vehicle has become a connected and autonomous vehicle. However, there is a lack of a common, decentralized and secure internet-of-vehicle (IoV) platform that can facilitate all of these applications not only for a few individuals but also for other key service providers by recording, analyzing and evaluating big data from vehicle sensors and ECUs.

*C. Motivation, Contribution and Organisation*

Vehicle Health Records are confidential information and require a high level of security and privacy to be dealt with. The access to the VHR can be logged onto the blockchain increasing the transparency on data access made by different stakeholders of the systems. Logging the data onto the public blockchain will also make data audability easier. The VHR hashes can be encoded and stored on distributed nodes of the Blockchain. Blockchain-based VHR framework can facilitate security, privacy, immutability, and required levels of accessibility, control, and transparency. The VHR (a portion of it or full) in the Blockchain can be automatically sent to the ITS service providers, insurance companies, manufacturers, concerned mechanics, and other parties securely.

Our twofold contributions are proposing a novel blockchain-based VHR architecture to enable various useful services in IoV and provide a summary of how it can be implemented. This paper presents an approach of capitalizing VHR not only for real-time health monitoring of the vehicle but also for various other important services that a vehicle may need in a secured manner. One can extend our framework to solve many other real-life problems of IoV and its associated services.

The rest of the paper is organized as follows: Section II contains the details of proposed system architecture. We discuss our goals in Section III and finally Section IV presents conclusion and future work.

## II. Overview of the V-CARE Architecture

In this section, we discuss the system model and working principles of the proposed V-CARE architecture shown in Fig. 3. We consider the flourished stage of IoV, with RSU, fog nodes, cloud, connected and smart vehicles, and other digitally-driven services.

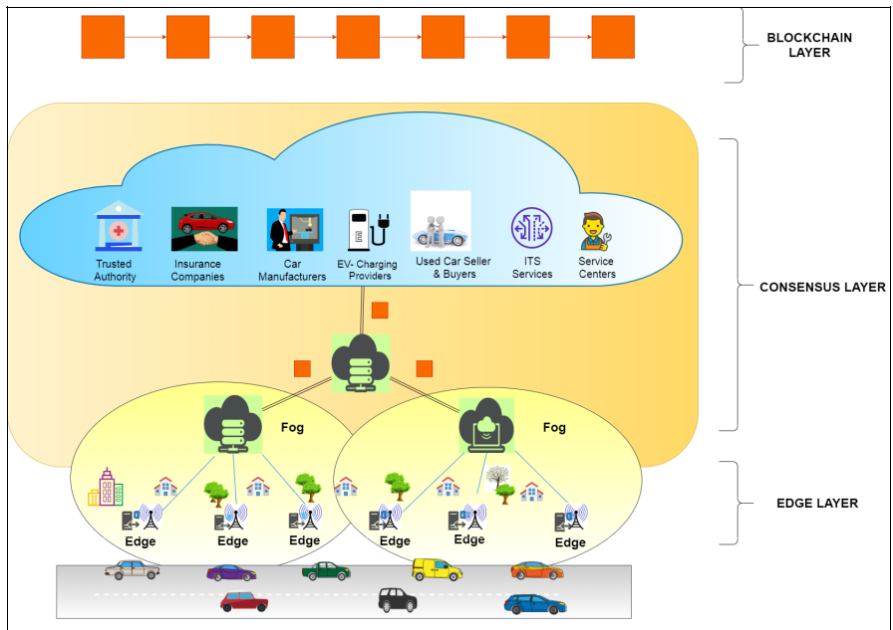

Fig. 3: Proposed V-CARE Framework

### A. Entities involved

*1) Vehicles:* The vehicles are equipped with sensors, ECUs, OBUs, data loggers (CAN, OBD-II, LIN, etc.), com-munication (V2V and V2X), storage, and computing facilities.

*2) RSUs:* The roadside units are deployed to support V2I/I2V connectivity. The connectivity is secure and can be used for data download/upload, over-the-air updates, etc [10]. The front-haul is wireless, and the back-haul is a wired backbone network and also securely connects the fog nodes.

*3) Regional Fog computing nodes:* The area selected can be segregated into small regions depending on traffic loads and peak hour analysis of entering and leaving vehicles. We propose fog nodes (FNs) in those regions with connectivity to one or more RSUs.

*4) Trusted Authority:* The trusted authority (TA) is a central entity using a PKI-based authentication and respon-sible for the registration of the vehicles, RSUs, regional fog nodes, and service providers. The TA issues cryptographic materials and security mechanisms to be employed in IoV. The TA maintains a global contract, Registry Contract (RC), that maps registered identity strings to Ethereum addresses, i.e., addresses on the blockchain. The entries available in RC link each entity to their activity contract.

*5) Service Providers:* As we have mentioned, these entities could be manufacturers, insurers, ITS administration, and other service providers. These service providers use a cloud-based system for data storage, processing, analysis, and assessment.

### B. Blockchain Related Terminologies

Blockchain has emerged as a disrupting technology over the last few years and has attracted interests from both industries as well as from the academic research community. With its inherent features of decentralization, fault-tolerance, non-reputability, high availability, high security, and reducing the need for trust, the technology is viewed as one of the contenders for solving some of the crucial challenges in the field of Finance, Supply chains, IoT and cloud computing.

*1) Fundamentals of Blockchain::* Blockchain gained global attention with the introduction of Bitcoin[11], a peer-to-peer payment system based upon cryptographic primitives. A blockchain is made up of a number of blocks where each block contains one or more number of transactions that are digitally signed by the users executing them with Public Key Scheme. These blocks are chained with each other with the help of cryptographic hashes, where each block in the chain points to its previous block. A blockchain can be conceptualized as a single linked list where one can traverse through the entire historical records following the links to the date of initialization of the chain.

Blockchain being a distributed system, requires a mechanism to make the nodes in the system agree to a consistent system state at a particular instance of time [12]. This is also sometimes known as reaching a consensus. There exist three types of nodes in a blockchain network:
- Light Node: These are low-end resource-constraint devices such as a smartphone.
- Full Node: These are the devices having sufficient storage capability but may not have a well computational facility.
- Miner Nodes: These are the devices having sufficient storage capability and high computational resources.

The responsibility of maintaining the blockchain network and making the nodes in the network reach a consensus state lies with the miner nodes. Miner nodes compete among themselves in order to solve a cryptographic challenge thrown out to them by the system.

*2) Smart Contracts to be used in V-CARE:*

*a) Registry Contract (RC):* The Trusted Authority (TA) is responsible for deploying the Registry Contracts (RCs). RCs provide information such as the date of registration of a vehicle, date of expiry of the license, addresses of valid cloud service providers (CSPs), etc. Vehicle owners, as well as the CSPs, need to approach the TA for the registration. The registration details are bound to the pseudonymous identifiers or the addresses, which corresponds to the identity of the entity in the real world. The mapping details of the entities to their pseudonymous identity is only known to the TA.

*b) Vehicle-Service Provider Relationship Contract (VSRC):* It defines how data will be managed and accessed by various entities. It contains an array of data pointers and corresponding access permissions that identify the records available at service providers. It is issued for pairwise data interaction between the vehicle and the service providers. The interaction happens through the execution of a query string available with data pointers. After the execution, it will return a subset of vehicle data. The query string is attached to the hash of the returned data to ensure integrity. Each such string specifies a portion of the vehicle's data that any service provider is allowed to access. VSRC also defines the terms of use of the data accessed by an entity.

*c) Activity Contract (AC):* VSRCs are linked through references available in ACs, so that vehicle records can be unified as a whole as and when required. Thus, it maintains a list of references to VSRCs, representing the previous and current interactions among various entities of the system. The ACs also facilitate essential backup, restoring, and notification functionalities. Vehicles can leave several times; however, get back to their history by downloading the latest blockchain after rejoining the system. Thus, ACs also maintain the blockchain logs.

*3) Mining and Transaction Types::* The blockchain entities in V-CARE reach the consensus by performing the mining procedure. Proof-of-Work (PoW) can be considered as the consensus mechanism in V-CARE as the network is supposed to be a public network. PoW outperforms other consensus mechanisms in terms of scalability in the number of nodes for public blockchains. Each miner participates in the mining procedure by packaging the incoming transactions in a block and solving a cryptographically hard problem that includes finding a nonce value satisfying some pre-conditions.

In V-CARE we propose to have two types of transactions:
- *Single Signed Transactions:* These transactions are the general type of blockchain transactions where a transaction needs to be signed by the initiator of the transaction. In V-CARE access requests made by a stakeholder such as an insurance company to a CSP of vehicle is a single signed transaction.
- *Multi-signature Transactions:* These are the transactions that are required to be signed by multiple identities where the number of identities must be at least more than one. Multi-signature transactions are much beneficial when consent from multiple parties is required. In V-CARE transactions resulting against data upload from the vehicle to the CSP is a multi-signature transaction which requires signatures from both the vehicle as well as the CSP to be valid.

*C. Working Principles*

   *1) Phase I: Registration:* All the vehicles (V1...Vn) first register themselves to the RC via the TA in order to join the blockchain-based VHR network. All service providers also need to register themselves in RC. Upon successful registration the entity is provided with an address and a private key corresponding to the address. The entity is uniquely identified on the blockchain network with the assigned address identifier.

   *2) Phase II: Data Logging, and Record at Vehicle:* The raw data generated from various ECUs, and OBD-II of the vehicle are logged (lossless logging) locally in the SDcard of onboard storage of the vehicle.

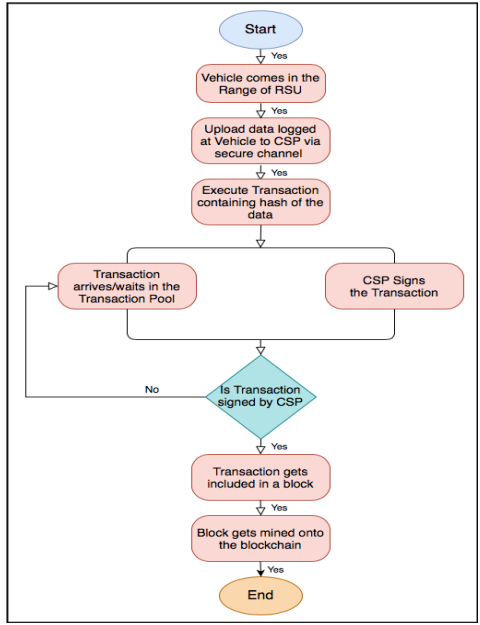 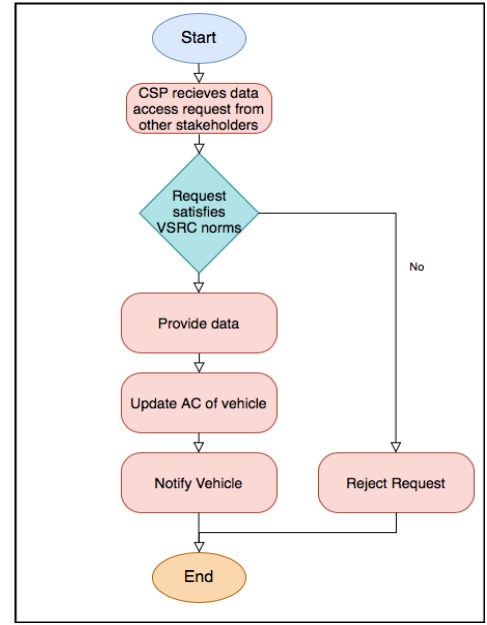

(a) Transaction processing in V-CARE.  (b) Access management in V-CARE

Fig. 4: Transaction Management and Access Management in V-CARE.

   *3) Phase III: Data Transfer to Fog node via RSUs:* The data recorded and stored on the vehicles are uploaded to the fog nodes via the RSU when the vehicles are in the RSU range. Once the backup is taken at the edge nodes, that data from vehicle storage can be purged to utilize the storage for local recording. The deeper analysis of the data is done at the cloud level by the various service providers and stakeholders of the system.

   *4) Phase IV: System Initiation:* The V-CARE is initialized by the TA with the deployment of infrastructures such as the RSU (Edge nodes) and the fog nodes. The TA sets up the blockchain network and does the deployment of smart contracts onto the chain.

   *5) Phase V: Consensus and Processing:* Once a vehicle comes under the range of an RSU, it initiates its data upload mechanism to the cloud via the Fog Nodes. The data upload is made through a secure channel. For each block of data being uploaded, the vehicle also initiates a multi-signature blockchain transaction corresponding to the data upload containing details such as a hash of the data block, timestamp, address of the intended cloud service provider. Each cloud service provider which also acts as a miner node responsible for maintaining the blockchain. Upon receiving a transaction, they check the credibility of the transaction first; then, they check whether any data block is present in their system with the corresponding transaction data hash. If yes, the CSP also signs the transaction. Once, both the signatures are acquired the miners bundle the transaction into a block and compete among themselves in order to get the block onto the blockchain. The transaction processing and access management flow chart is shown in Fig. 4 (a) and Fig. 4 (b), respectively. In V-CARE, the fog nodes deployed by the TA and the CSPs primarily act as the miner nodes, however other stakeholders of the ecosystem can also join the mining procedure by deploying their own mining equipment to make the blockchain much more decentralized.

*6) Phase VI: Access Control:* Once, a CSP receives a data access request from another stakeholder for a vehicle whose data management is associated with the CSP, the CSP checks for the satisfaction of conditions laid down by the VSRC of the vehicle. If the request satisfies the VSRC norms, access to the data is provided to the requester, and the vehicle's AC is also updated, marking a successful data access attempt.

*7) Phase VII: Service Delivery:* Service providers perform different analyses on the data that they receive from the CSP, such as the manufacturer of the vehicle analyzing the OBD-II and ECUs specific data. In contrast, the insurance company might be more interested in analyzing the driving behavior data. Based on the analysis, the critical information will be forwarded immediately to drivers as alerts, and warnings. The other important information can be offloaded to the regional fog, which will be pushed to the visiting edge nodes for better service delivery.

## III. Discussion

We proposed a blockchain-based VHR system called V-CARE that leverages a flourished multi-tier [13], [14] IoV framework (Cloud, Fog, and Edge nodes) for better service delivery to vehicles. Our main concern was how to utilize the power of blockchain to design a secure VHR that fulfill the security requirements such as integrity, authenticity, non-repudiation, availability, access control, etc. Our focus was on how the scope of VHR can be broadened to facilitate a wide variety of services in IoV. With the use of contracts, various activities by all the entities can be logged and recorded in a consistent manner in the blockchain. The data, once recorded, can't be altered by any parties. Vehicles are highly dynamic and frequently leave and join the network. Once they rejoin the system, the proposed architecture can help them upload data, access results, and other activities in a secure and fast manner. The proposed framework also facilitates backup, restore, and warning systems through activity contracts.

## IV. Conclusion and Future Work

In this paper, we presented a concept of a novel decentralized VHR system that leverages blockchain technology for maintaining vehicle health records, which can enable various types of services in IoV. We discussed how such a decentralized approach could help to handle VHRs securely and provide a wide range of services by accessing them. In the near future, we look forward to implementing the system on available blockchain platforms to demonstrate the feasibility. We will try to integrate quantum resilience multi-factor authentication and attribute-based access control for data access and sharing across different entities. We will provide details of VHR data update policies, such as how data will be added and deleted. Our work can be extended to integrate more services and solve other challenges of IoV.